\documentclass[aps,prd,onecolumn,groupedaddress,showpacs,nofootinbib,amssymb]{revtex4}
\usepackage[T1]{fontenc}
\usepackage[latin1]{inputenc}
\usepackage{graphicx}
\usepackage[english]{babel}
\usepackage{amsmath}
\usepackage{amssymb}
\usepackage{amsfonts}

\begin{document}

\def\pp{{\, \mid \hskip -1.5mm =}}
\def\cL{{\cal L}}
\def\be{\begin{equation}}
\def\ee{\end{equation}}
\def\bea{\begin{eqnarray}}
\def\eea{\end{eqnarray}}
\def\beq{\begin{eqnarray}}
\def\eeq{\end{eqnarray}}
\def\tr{{\rm tr}\, }
\def\nn{\nonumber \\}
\def\e{{\rm e}}

\title{Is the future universe singular: Dark Matter versus modified gravity?}

\author{Shin'ichi Nojiri$^1$ and Sergei D. Odintsov$^{2,3}$\footnote{
Also at Center of Theor. Physics, TSPU, Tomsk}}
\affiliation{
$^1$Department of Physics, Nagoya University, Nagoya 464-8602, Japan \\
$^2$Instituci\`{o} Catalana de Recerca i Estudis Avan\c{c}ats (ICREA),
Barcelona \\
$^3$ Institut de Ciencies de l'Espai (IEEC-CSIC),
Campus UAB, Facultat de Ciencies, Torre C5-Par-2a pl, E-08193 Bellaterra (Barcelona), Spain}

\begin{abstract}

The fundamental problem of the occurrence/removal of finite-time future singularity in 
the universe evolution  for coupled dark energy (DE) is addressed. It is demonstrated the existence 
of the (instable or local minimum) de Sitter space solution which may cure the Type II or Type IV 
future singularity for DE coupled with DM as the result of tuning the  initial conditions. In case 
of phantom DE, the corresponding coupling may help to resolve the coincidence problem but not 
the Big Rip (Type I) singularity issue.
We show that modified gravity of special form or inhomogeneous DE fluid may offer the universal 
scenario to cure the Type I,II,III or IV future singularity of coupled (fluid or scalar) DE evolution.

\end{abstract}


\maketitle
\section{Introduction}

The singularity issue has the fundamental importance in the modern cosmology.
The early universe may appear totally different what depends on the presence/absence 
of the initial singularity. There are various points of view on this issue.
Generally speaking, the singularity problem may be properly understood/resolved only 
in full quantum gravity theory which does not exist so far.

With the discovery of the late-time acceleration the singularity problem became even more important. 
The reason is that observational data favour the $\Lambda$CDM cosmology with the 
equation of state (EoS) parameter being very close to $-1$. This means that phantom/quintessence 
Dark Energy (DE) models with effective EoS parameter $w$ approximately equal to $-1$ are not excluded. 
Unfortunately, it was discovered \cite{Nojiri:2005sx} 
that many of such DEs may lead to one of four different finite-time, future singularities 
in the universe evolution. 
Definitely, the presence of finite-time, future singularity may cause various problems/instabilities 
in the current black holes and stellar astrophysics. Hence, it is very interesting to understand 
if any natural scenario to cure such singularities exist. The purpose of present work is to consider 
the classical scenario of the future singularity removal for coupled DE. In the next section 
we consider phantom DE coupled with Dark Matter(DM). It is demonstrated the existence of instable de Sitter 
solution which may solve the coincidence problem but does not cure the future singularity. 
Third section is devoted to the consideration of DE fluid which may develop all four types of 
future singularity. It is shown that its coupling with DM may cure Type II and Type IV singularities 
but not Big Rip (Type I) and Type III ones.
In fourth section it is demonstrated that special models of modified gravity or of the inhomogeneous 
EoS fluids \cite{Nojiri:2005sr} may offer the universal scenario to cure the future singularity of 
any type. This is demonstrated for fluid or scalar-tensor singular DE.
Some outlook is given in the conclusion.

\section{Decaying Phantom Dark Energy: the Solution to the Coincidence Problem}

Let us remind several simple facts about coupled phantom DE.
The DE conservation law is given by
\be
\label{OI}
\dot \rho_\mathrm{DE} + 3H \left(1 + w \right)\rho_\mathrm{DE} = 0\, ,
\ee
while the first FRW equation
\be
\label{OII}
\frac{3}{\kappa^2} H^2 = \rho_\mathrm{DE}\, ,
\ee
admits a solution ( $w$ is less than $-1$)
\be
\label{OIII} 
H = \frac{\frac{2}{3\left( 1 + w \right)}}{t_s - t}\, ,
\ee
which has finite-time future singularity, called Big Rip, at $t=t_s$
(for number of earlier works studying singular phantom DE era, see \cite{ref5,Elizalde:2004mq}).

We now consider the model where phantom DE couples with dark matter.
The conservation law is
\be
\label{I}
\dot \rho_\mathrm{DE} + 3H \left(1 + w \right)\rho_\mathrm{DE} = - Q \rho_\mathrm{DE}\, ,
\quad \dot \rho_\mathrm{DM} + 3H \rho_\mathrm{DM} = Q \rho_\mathrm{DE}\, .
\ee
Here $Q$ is assumed to be a constant. The first equation can be solved as
\be
\label{II}
\rho_\mathrm{DE} = \rho_\mathrm{DE(0)} a^{-3\left(1 + w \right)} \e^{-Qt}\, .
\ee
Here $\rho_\mathrm{DE(0)}$ is a constant of the integration. Then the
second equation (\ref{I}) can be solved as
\be
\label{III}
\rho_\mathrm{DM} = Q a(t)^{-3} \int^t dt' \rho_\mathrm{DE(0)} a^{- 3 w} \e^{-Qt}\, .
\ee
The second FRW equation is now given by
\be
\label{IV}
 - \frac{1}{\kappa^2}\left(2\dot H + 3H^2\right) = p = w \rho_\mathrm{DE}
= w \rho_\mathrm{DE(0)} a^{-3\left(1 + w \right)} \e^{-Qt}\, .
\ee
An exact solution of (\ref{IV}) is de Sitter space
\be
\label{V}
a(t) = a_0 \e^{- \frac{Q}{3\left(1 + w \right)} t}\quad
\left(H= - \frac{Q}{3\left(1 + w \right)} \right)\, ,
\ee
where $a_0$ is given by
\be
\label{VI}
 - \frac{3}{\kappa^2} \left(\frac{Q}{3\left(1 + w \right)}\right)^2
= w \rho_\mathrm{DE(0)} a_0^{-3\left(1 + w \right)}\ .
\ee
Note that we are considering phantom DE with $w<-1$ and therefore
$H$ (\ref{V}) is positive and Eq.(\ref{VI}) has a real solution.

Hence, the coupling of the dark matter with the phantom DE gives
de Sitter solution instead of Big Rip approaching solution (\ref{OIII}).
This does not always mean that Big Rip singularity could be avoided
but gives a possibility that the universe could evolve to de Sitter universe instead of Big Rip.

One may identify the Hubble rate $H$ with the present value of the Hubble rate
$ H = - \frac{Q}{3\left(1 + w \right)} = H_0 \sim 10^{-33}\, \mathrm{eV}$.
Eq.(\ref{II}) shows that DE density is a constant
\be
\label{VII}
\rho_\mathrm{DE} = \rho_\mathrm{DE(0)} a_0^{-3\left(1 + w \right)} \, .
\ee
Then Eq.(\ref{III}) can be integrated as
\be
\label{VIII}
\rho_\mathrm{DM} = \rho_\mathrm{DM\, 0} a^{-3}
 - \left( 1+ w \right) \rho_\mathrm{DE(0)} a_0^{-3\left(1 + w \right)}\, .
\ee
Here $\rho_\mathrm{DM\, 0}$ is a constant of the integration but the first FRW equation
\be
\label{IX}
\frac{3}{\kappa^2}H^2 = \rho_\mathrm{DM} + \rho_\mathrm{DE}\ ,
\ee
shows $\rho_\mathrm{DM\, 0}=0$ and therefore the dark matter density $\rho_\mathrm{DM}$ is
also constant:
\be
\label{X}
\rho_\mathrm{DM} =
 - \left( 1+ w \right) \rho_\mathrm{DE(0)} a_0^{-3\left(1 + w \right)}
= - \left( 1+ w \right) \rho_\mathrm{DE}\, .
\ee
Then if the de Sitter solution (\ref{V}) is an attractor, by choosing
\be
\label{XI}
 - \left( 1+ w \right) \sim \frac{1}{3}\ ,\quad \mbox{i.e.}\quad w \sim - \frac{4}{3}\ ,
\ee
the coincidence problem could be solved. That is, even if we start with a wide range of
the initial conditions, the solution approaches to the de Sitter solution, where the ratio
of DE and DM is approximately $1/3$ almost independent from the initial
condition.

If DM does not couple with DM, the densities behave as
$\rho_\mathrm{DM} \sim a^{-3}$ and $\rho_\mathrm{DE} \sim a^{-3\left( 1 + w \right)}$,
that is, DM density decreases but the DE density increases when
the universe expands. This requires the fine-tuning for the initial condition of the
densities so that the density of DM has almost same order with that
of DE. This is so-called coincidence problem.

In order to investigate if the de Sitter solution (\ref{V}) is an attractor or not,
we consider the perturbation as
\be
\label{XII}
a = a_0 \e^{- \frac{Q}{3\left(1 + w \right)} t + \Delta(t)}\ .
\ee
Here $\Delta(t)$ is assumed to be small.
The second FRW equation (\ref{IV}) gives
\be
\label{XIII}
 - \frac{1}{\kappa^2}\left(2\ddot \Delta - \frac{2Q}{\left(1 + w \right)}\dot\Delta\right)
= -3 \left(1 + w \right)w \rho_\mathrm{DE(0)} a_0^{-3\left(1 + w \right)} \Delta
= \frac{3\left(1 + w \right)}{\kappa^2} \left(\frac{Q}{3\left(1 + w \right)}\right)^2 \Delta \, ,
\ee
which is very simple linear differential equation with constant coefficient.
Here Eq.(\ref{VI}) is used.
By assuming $\Delta = \e^{\lambda t}$, we find
\be
\label{XIIIB}
0= \lambda^2 - \frac{Q}{1 + w} \lambda
+ 3\left(1 + w \right) \left(\frac{Q}{3\left(1 + w \right)}\right)^2 \, ,
\ee
that is
\be
\label{XIV}
\lambda = \lambda_\pm \equiv \frac{Q}{2\left(1 + w \right)} \pm \frac{1}{2}
\left\{ \left( \frac{Q}{1 + w} \right)^2
 - \frac{4\left(1 + w \right)}{3} \left(\frac{Q}{\left(1 + w \right)}\right)^2 \right\}^\frac{1}{2}\ .
\ee
Then $\lambda_- < 0$ but $\lambda_+>0$ and therefore the de Sitter solution
(\ref{V}) is not stable. Since $\lambda_- < 0$, however, the solution is saddle point and therefore
if we choose the appropriate initial condition (not by fine-tuning), there is a solution approaching
to the saddle point de Sitter solution (\ref{V}).
Although we need to study the global structure of the space of the solutions, such the initial condition
could correspond to the direction towards to $\lambda_-$.
Hence, generally speaking, the coupling of DE with DM does not prohibit the existence
of the (Big Rip) singular solution.
Then if the singularity corresponds to the stable solution, since the de Sitter space is not
completely stable, although there is a solution approaching to the de Sitter solution, the solution
will finally evolve to the singular solution.
Hence, the appropriate choice of initial conditions may help to realize
the non-singular de Sitter cosmology which also solves the coincidence problem.
At least, one may expect that the future singularity occurs at more late times than without coupled DM.

\section{Coupling of Dark Energy with Dark Matter: Singularity avoidance?}

Let us consider the simple example of perfect fluid with the following
equation of state (EoS) \cite{Barrow:1990vx}:
\be
\label{EOS1}
p = - \rho + A \rho^\alpha\ ,
\ee
with constant $A$ and $\alpha$.
We work in the spatially flat FRW space-time :
$ds^2 = - dt^2 + a(t)^2 \sum_{i=1,2,3} \left(dx^i\right)^2$.
Then the Hubble rate is found to be
\be
\label{EOS6}
H = \left\{ \begin{array}{ll}
\frac{\frac{3}{2}A}{t}\ ,\quad & \mbox{when}\ \alpha = 1\ ,\quad A>0 \\
\frac{-\frac{3}{2}A}{t_0 - t}\ ,\quad & \mbox{when}\ \alpha = 1\ ,\quad A<0 \\
B \e^{- \frac{\sqrt{3}\kappa A t}{2}}\ ,\quad &
\mbox{when}\ \alpha = \frac{1}{2}\ ,\quad A<0 \\
\begin{array}{l}
C t^{1/(1-2\alpha)} \\
\mbox{or}\ \tilde C \left(t_0 - t\right)^{1/(1-2\alpha)}
\end{array} & \mbox{when}\ \alpha \neq 1,\ \frac{1}{2}
\end{array} \right.
\ee
Here $B$, $C$, and $\tilde C$ are positive constants.
Now one can describe the future, finite-time singularities of the universe
filled with above dark fluid for different choices of theory parameters (see ref.\cite{Nojiri:2009uu}). 
For more complicated DEs leading to all four types of future singularity, 
see \cite{Nojiri:2005sr, singularity}.
When $\alpha<0$, there occurs Type II or sudden future singularity \cite{barrow,singularity}.

When $0<\alpha<1/2$ and $1/(1-2\alpha)$ is not an integer, there occurs
Type IV singularity.
When $\alpha=0$, there is no any singularity.
When $1/2<\alpha<1$ or $\alpha=1$ and $A<0$, there appears Type I or Big Rip type
singularity.
When $\alpha>1$, there occurs Type III singularity
(for classification of all four types of future singularity, see \cite{Nojiri:2005sx}).

In case of Type II singularity, where $\alpha<0$, $H$ vanishes as
$H \sim \left(t_0 - t\right)^{1/(1-2\alpha)}$ when $t\to t_0$
and therefore we find $\rho$ vanishes as it follows from
the FRW equation : $\left(3/\kappa^2\right)H^2 = \rho$. Then, near the singularity, the EoS
(\ref{EOS1}) is reduced to
\be
\label{EoSt1}
p \sim A\rho^{\alpha}\ .
\ee
On the other hand, in case of Type I singularity, where $1/2<\alpha<1$
or $\alpha=1$ and $A<0$, $H$ and therefore $\rho$ diverges when $t\to t_0$. Then the EoS (\ref{EOS1})
reduces to
\be
\label{EoSt2}
p \sim - \rho\ \mbox{or}\ p\sim -(1-A)\rho\ .
\ee
In case of Type III singularity, where $\alpha>1$,
$H$ and $\rho$ diverge when $t\to t_0$ and therefore the EoS
(\ref{EOS1})
reduces to
\be
\label{EoSt3}
p \sim A \rho^\alpha\ .
\ee

We now consider DE (\ref{EOS1}) coupled with DM as in
(\ref{I}).
Then the conservation law is given by
\be
\label{AI}
\dot \rho_\mathrm{DE} + 3H A \rho_\mathrm{DE}^\alpha = - Q \rho_\mathrm{DE}\, ,
\quad \dot \rho_\mathrm{DM} + 3H \rho_\mathrm{DM} = Q \rho_\mathrm{DE}\, .
\ee
The solution of (\ref{AI}) is
\be
\label{AII}
\rho_\mathrm{DE}(t) = \e^{-Qt} \left( - 3A\left(1-\alpha\right)
\int^t dt' H(t') \e^{\left(1-\alpha\right) Qt'}
\right)^{\frac{1}{1-\alpha}}\, , \quad
\rho_\mathrm{DM} = Q a(t)^{-3} \int^t dt' a(t')^3 \rho_\mathrm{DE}(t')\, .
\ee
Then if $A<0$ (and $Q>0$), there is a de Sitter solution $H=H_0$ with
a constant $H_0$, which is given by solving
\be
\label{AIII}
\frac{3}{\kappa^2} H_0^2 = \left( 1 + \frac{Q}{3H_0}\right)
\left( - \frac{3A H_0}{Q}\right)^{\frac{1}{1-\alpha}}\, ,
\ee
and $\rho_\mathrm{DE}$ and $\rho_\mathrm{DM}$ are constants
\be
\label{AIV}
\rho_\mathrm{DE}
= \left( - \frac{3A H_0}{Q}\right)^{\frac{1}{1-\alpha}}\, ,\quad
\rho_\mathrm{DM}
= \frac{Q}{3H_0}\left( - \frac{3A H_0}{Q}\right)^{\frac{1}{1-\alpha}}
\, .
\ee
This demonstrates that if $H_0\sim Q$, we find $\rho_\mathrm{DM}/\rho_\mathrm{DE} \sim 1/3$ and the
coincidence problem may be solved. If $H_0= Q$, Eq.(\ref{AIII}) determines the value of $A$:
\be
\label{AV}
A = - \frac{1}{3}\left( \frac{9}{4\kappa^2} H_0^2 \right)^{1-\alpha}\, .
\ee

One now investigates the (in)stability of the de Sitter solution $H=H_0$ by putting $H=H_0 + \delta H$.
The perturbation of energy-density is
\be
\label{AVI}
\delta \rho_\mathrm{DE} = -3A \left( - \frac{3AH_0}{Q}\right)^{ \frac{\alpha}{1-\alpha}}
\e^{-\left(1 - \alpha\right) Qt}\int^t dt' \delta H (t')
\e^{\left(1 - \alpha\right) Qt'}\, .
\ee
The second FRW equation becomes
\be
\label{AVII}
 - \frac{1}{\kappa^2}\left(\delta \dot H + 6 H_0 \delta H\right) \e^{\left(1 - \alpha\right) Qt'}
= 3A \left( 1 + \frac{\alpha Q}{3H_0} \right)
\left( - \frac{3AH_0}{Q}\right)^{ \frac{\alpha}{1-\alpha}}\int^t dt' \delta H (t')
\e^{\left(1 - \alpha\right) Qt'}\, .
\ee
By differentiating the both sides of (\ref{AVII}), we find
\bea
\label{AVIII}
0 &=& \delta \ddot H + \left\{ 6H_0 + \left(1 - \alpha\right) Q \right\} \delta \dot H
+ \left\{ 6H_0 \left(1 - \alpha\right) Q + 3A \kappa^2 \left( 1 + \frac{\alpha Q}{3H_0} \right)
\left( - \frac{3AH_0}{Q}\right)^{ \frac{\alpha}{1-\alpha}} \right\} \delta H \nn
&=& \delta \ddot H + \left\{ 6H_0 + \left(1 - \alpha\right) Q \right\} \delta \dot H
+ 3H_0 \left( 1 - 2\alpha \right) Q \delta H \, .
\eea
In the second equality, Eq.(\ref{AIII}) is used.
Assuming $\delta H \propto \e^{\lambda t}$, one gets
\be
\label{AVIIIA}
0 = \lambda^2 + \left\{ 6H_0 + \left(1 - \alpha\right) Q \right\} \lambda
+ 3H_0 \left( 1 - 2\alpha \right) Q \, ,
\ee
whose solution is given by
\be
\label{AIX}
\lambda = \lambda_\pm \equiv - \frac{6H_0 + \left(1 - \alpha\right) Q}{2}
\pm \frac{1}{2} \left\{ \left\{ 6H_0 + \left(1 - \alpha\right) Q \right\}^2
 - 12 H_0 \left( 1 - 2\alpha \right) Q \right\}^{\frac{1}{2}}\ .
\ee
Then since $H_0$, $Q>0$, if
\be
\label{AX}
\alpha < \frac{1}{2}\, ,
\ee
both of $\lambda_\pm$ are real and negative if
\be
\label{AXI}
D= \left(6H_0 + \left(1 - \alpha\right) Q\right)^2 - 12H_0 \left( 1 - 2\alpha \right) Q > 0\, ,
\ee
or complex but the real part is negative if
\be
\label{AXII}
D= \left(6H_0 + \left(1 - \alpha\right) Q\right)^2 - 12H_0 \left( 1 - 2\alpha \right) Q < 0\, .
\ee
Then as long as $\alpha < \frac{1}{2}$ (\ref{AX}), the de Sitter solution is stable
and therefore the singularity could be avoided.
On the other hand, if
\be
\label{AXIII}
6H_0 + \left(1 - \alpha\right) Q > 0\, ,\quad
3H_0 \left( 1 - 2\alpha \right) Q < 0\, ,
\ee
that is
\be
\label{AIXX}
1 + \frac{6H_0}{Q} > \alpha > \frac{1}{2}\, ,
\ee
we find $\lambda_+ >0$ and $\lambda_- <0$, as in the case of the previous section. In fact,
the previous section corresponds to $\alpha = 1$ case.
If
\be
\label{AXX}
6H_0 + \left(1 - \alpha\right) Q < 0\, ,\quad
3H_0 \left( 1 - 2\alpha \right) Q < 0\, ,
\ee
that is
\be
\label{AXXI}
\alpha > 1 + \frac{6H_0}{Q} \, ,
\ee
we find $\lambda_\pm >0$ and the de Sitter solution is totally unstable.

The existence of the singular solution itself, even in the presence of coupled dark matter,
can be confirmed by substituting the singular solution
without DM into the equations.

Hence,
Type II singularity, where $\alpha<0$, and
Type IV singularity, where $0<\alpha<1/2$ and $1/(1-2\alpha)$ is not an integer,
could be cured by the coupling of DE with DM.
As we mentioned, even if there is a coupling of DM with DE,
there could be a singular solution. The de Sitter solution with $\alpha < \frac{1}{2}$ is,
however, at least a local minimum. Then if universe started with an appropriate initial
condition, the universe evolves into the de Sitter one (asymptotically de Sitter universe).
Type I (Big Rip) singularity, where $1/2<\alpha<1$, and Type III singularity, where $\alpha>1$,
could not be removed by the coupling of DM with DE. Even if a solution goes
near the de Sitter point, the solution could evolve into the singular solution
if the singular solution is stable.
Nevertheless, the fact of avoidance of some future singularities due to coupling of DE
with DM looks quite promising.

\section{Modified gravity curing the singularity}

Since the coupling of DE with DM does not always remove the singularity, we now consider
what kind of the fluid could cure the future singularity. In case of the Big Rip singularity,
for example, the energy-density of DE diverges like
$\rho_\mathrm{DE} \sim 1/\left( t_s - t \right)^2 \sim R$ when $t\to t_s$.
Here $R$ is the scalar curvature. Then one is interesting in a fluid whose pressure
is positive (and energy-density is positive) and grows up more rapidly than DE pressure.
 There is no such a fluid with constant EoS parameter. However, one can consider the
pressure which is proportional to a power of the curvature, for example,
\be
\label{B1}
p_\mathrm{fluid} \propto R^{1 + \epsilon}\, ,
\ee
with $\epsilon>0$.
Then the total EoS parameter becomes greater than $-1$ for large curvature and
Big Rip does not occur.

This kind of inhomogeneous effective fluid \cite{Nojiri:2005sr} could be realized by quantum effects
(for instance, taking account of conformal anomaly)
or by modified gravity (for general review, see \cite{Nojiri:2006ri}).
As an example, we consider $F(R)$ gravity, where $F(R)=R + f(R)$ behaves as $f(R) \propto R^m$.

When $R$ is large, if $m>1$, the contribution from matter, DM and DE
could be neglected. There occurs the following solution:
\be
\label{JGRG17}
H \sim \frac{-\frac{(m-1)(2m-1)}{m-2}}{t}\, ,
\ee
which gives the following effective EoS parameter:
\be
\label{JGRG18}
w_{\rm eff} \equiv - 1 - \frac{2\dot H}{3H^2}
= - 1 - \frac{2(m-2)}{3(m-1)(2m-1)}\, .
\ee
which is greater than $-1$ if $2>m>1$ or $m<1/2$. In case of $m<1/2$, however, the Einstein-Hilbert
term $R$ and/or the dark matter could dominate over $f(R)$ term and we
do not consider this case. In case of $2>m>1$, since $w_{\rm eff}>-1$, there does not occur the Big Rip
singularity or any kind of the future singularity. We should note that there occurs the Big Rip
singularity when $m>2$. In case of $m=2$,
as Eq.(\ref{JGRG17}) tells, the power law solution like Big Rip singularity is prohibited.
We can find that there appear
(asymptotic) de Sitter solution, instead of the power law solution as in (\ref{JGRG17}),
which is consistent with (\ref{JGRG18}), and therefore the singularity
is prohibited.
Therefore, the introduction of special form of $f(R)$-term 
prevents the future singularity. Note that as other DEs, the modified gravity itself may lead to 
all four possible future singularities \cite{singularity} which may be cured by $R^2$-term 
\cite{Abdalla:2004sw, singularity} (for related discussion of curing the Type II singularity in 
special modified gravity by $R^2$ term, see \cite{other}).

On the other hand, if we choose
\be
\label{B2}
m = m_{\pm} = \frac{7 \pm \sqrt{73}}{2}\, ,
\ee
we find $w_{\rm eff}$ vanishes, which corresponds to the dust-like dark matter or usual matter.

Let us consider the following example:
\be
\label{B3}
F(R) \sim f_+ R^{m_0} + f_- R^{m_-}\, .
\ee
If we choose $m_0$ to be $2>m_0>1$, the first term will prevent the singularity when
the curvature is large and the second term
might behave as dark matter when the curvature is small since $m_-<0$.
In (\ref{JGRG18}), if $m>2$, there could occur Big Rip type singularity. In order that $H>0$ in
(\ref{JGRG17}), one may assume $t<0$ at present universe, or equivalently shift $t$
as $t\to t - t_0$ and rewrite (\ref{JGRG17}) as
\be
\label{B4}
H \sim \frac{\frac{(m-1)(2m-1)}{m-2}}{t_0 - t}\, .
\ee
Then if $t<t_0$ in the present universe, $H$ is positive and there occurs the Big Rip singularity
at $t=t_0$.
On the other hand we may consider the scalar tensor-theory whose action is given by
\be
\label{B5}
S_\mathrm{ST} = \int dx^4 \sqrt{-g} \left[ \frac{R}{2\kappa^2} - \frac{1}{2}
\partial_\mu \phi \partial^\mu \phi - V_0 \e^{-\frac{2\phi}{\phi_0}} \right] \, ,
\ee
with constants $V_0$ and $\phi_0$.
The action (\ref{B5}) admits the following solution
\be
\label{B6}
\phi=\phi_0\ln \left|\frac{t}{t_1}\right|\, ,\quad
H=\frac{\kappa^2\phi_0^2}{4 t}\, , \quad
t_1^2 \equiv -\frac{
\phi_0^2 \left(1 - \frac{3 \kappa^2\phi_0^2}{4}\right)}{2V_0}\, .
\ee
Then if
\be
\label{B7}
\kappa^2\phi_0^2 = \frac{8}{3}\, ,
\ee
we find $w_{\rm eff} \equiv - 1 - \frac{2\dot H}{3H^2} = 0$ and the case of (\ref{B7}) corresponds
to the dark matter.
We may consider the model, the scalar-tensor theory (\ref{B5}) coupled with $f(R) \propto R^m$-gravity
with $m>2$ like the Brans-Dicke theory as
\be
\label{B8}
S_\mathrm{ST} = \int dx^4 \sqrt{-g} \left[ \frac{R}{2\kappa^2} + f_0 R^m - \frac{1}{2}
\partial_\mu \phi \partial^\mu \phi - V_0 \e^{ -\frac{2\phi}{\phi_0}}
 - U_0 \e^{\frac{2\phi}{\phi_0}} R \right]\, ,
\ee
with a coupling $U_0$. The term with $U_0$ could express the interaction between gravity and the
scalar filed.
The action (\ref{B5}) admits the de Sitter solution as we see now. Assume the scalar field $\phi$ and
the curvatures is covariantly constant:
\be
\label{B9}
\phi = c\, ,\quad R=R_0\, ,\quad R_{\mu\nu} = \frac{1}{4} R_0 g_{\mu\nu}\, ,
\ee
with constants $c$ and $R_0$.
Then the equations corresponding to the Einstein one and the
scalar field have the following form:
\bea
\label{B10}
0 &=& \frac{R_0}{2\kappa^2} + (2-m) f_0 R_0^m - 2V_0 \e^{-\frac{2c}{\phi_0}}
 - U_0 \e^{\frac{2c}{\phi_0}} R_0 \, , \\
\label{B11}
0 &=& V_0 e^{-\frac{2c}{\phi_0}} - U_0 \e^{\frac{2c}{\phi_0}} R_0 \, .
\eea
Combining Eqs.(\ref{B10}) and (\ref{B11}), one obtains
\bea
\label{B12}
&& R = \frac{V_0}{U_0} \e^{- \frac{4c}{\phi_0}}\, ,\\
\label{B13}
&& \frac{V_0}{2\kappa^2 U_0}\e^{- \frac{2c}{\phi_0}}
+ \left( 2 - m \right) f_0 \left( \frac{V_0}{U_0} \right)^2 \e^{- \frac{(4m - 2)c}{\phi_0}}
= 3V_0\, .
\eea
If $V_0$, $U_0>0$, $f_0<0$, and $m>2$, the l.h.s. of (\ref{B13}) is positive and monotonically decreasing
function of $c$ and the l.h.s. vanishes in the limit of $\phi\to +\infty$ but positively diverges
in the limit of $\phi\to - \infty$. Therefore (\ref{B13}) has a unique solution with respect to $c$.
Then Eq.(\ref{B12}) shows $R$ is constant and positive, which expresses de Sitter universe.
We should note that when $U_0\to 0$, $R$ diverges and therefore there is no de Sitter solution without
the coupling $U_0$. 

Hence, we demonstrated that DE with inhomogeneous EoS corresponding to special form of modified gravity 
may easily cure the future singularity of any type. It indicates that if our universe does not like 
the future singularity, then modified gravity should play the role of DE.

\section{Discussion}

In summary, we considered DE model which may lead to all four types of future singularity in the 
late-time universe evolution. It is demonstrated that its coupling with DM may cure Type II and 
Type IV singularities (but not Big Rip and Type III) already on the classical level. It turns out that 
only modified gravity/inhomogeneous EoS DE may suggest the universal classical recipe to remove 
any of the known future singularities. This is shown for fluid DE as well as for scalar DE coupled 
with modified gravity. Moreover, there are viable 
non-singular modified gravities which describe the unification of the early-time inflation with 
late-time acceleration. Adding such non-singular modified gravity to singular DE model effectively 
removes the future singularity of any type as is described in ref.\cite{Nojiri:2009xw}. 
In this respect, the alternative gravity DE may seem to be more fundamental theory than 
more traditional (scalar, fluid, etc.) DE.

As final remark, one should stress that our consideration is totally classical. Nevertheless, it 
is expected that quantum gravity effects may play the significant role near to singularity. It is clear 
that such effects may contribute to the singularity occurrence/removal too. Unfortunately, 
due to the absence of complete quantum gravity only preliminary estimations may be done.
However, already the account of one-loop quantum gravity effects indicates to the possibility 
of the removal of future singularity \cite{Elizalde:2004mq}. This gives one more argument in favor 
of modified gravity as the possible universal regulator of future singularity. 

\section*{Acknowledgments}

The work by S.N. is supported in part by Global
COE Program of Nagoya University provided by the Japan Society
for the Promotion of Science (G07).
The work by S.D.O. is supported in part by MICINN (Spain) project
FIS2006-02842, by AGAUR (Generalitat de Catalunya), project 2009 SGR994
and by JSPS Visitor program.

\end{document}